\documentclass[prl,preprint,showpacs]{revtex4}

\usepackage[frenchb]{babel} 
\usepackage{epsfig}
\usepackage{amsmath}
\usepackage{graphicx}
\usepackage{bm}
\usepackage[T1]{fontenc}
\usepackage[latin1]{inputenc}
\usepackage{times}
\usepackage{ulem}
\usepackage{color}

\begin{document}

\title{Exploiting the Time-Reversal Operator for Adaptive Optics, Selective Focusing and Scattering Pattern Analysis}
\author{S.~M. Popoff}
\author{A. Aubry}\email{alexandre.aubry@espci.fr}
\author{G. Lerosey}\author{M. Fink}\author{A.~C. Boccara}\author{S. Gigan}

\affiliation{Institut Langevin, ESPCI ParisTech, CNRS UMR 7587, Universités Paris VI \& VII, INSERM, ESPCI, 10 rue Vauquelin, Paris, 75231, France} 

\date{\today}

\begin{abstract}
We report on the experimental measurement of the backscattering matrix of a weakly scattering medium in optics, composed of a few dispersed gold nanobeads. The DORT method (Decomposition of the Time Reversal Operator) is applied to this matrix and we demonstrate selective and efficient focusing  on individual scatterers, even through an aberrating layer. Moreover, we show that this approach  provides the decomposition of the scattering pattern of a single nanoparticle. These results open important perspectives for optical imaging, characterization and selective excitation of nanoparticles.
 \end{abstract}


\maketitle
The propagation of waves in inhomogeneous media is a fundamental problem which implies applications in many domains of wave physics, ranging from optics or acoustics to solid-state physics, seismology, medical imaging, or telecommunications. Conventional focusing and imaging techniques based on the Born approximation generally fail in aberrating or strongly scattering media due to the phase distorsions and the multiple scattering events undergone by the incident wavefront. Recent advances in light manipulation techniques have allowed great progresses in  optical focusing and imaging through complex media. Following pioneering works in ultrasound \cite{derode,derode2}, Vellekoop \textit{et al.} \cite{vellekoop2007focusing} showed how light can be  focused spatially through a strongly scattering medium by actively shaping the incident wavefront with a spatial light modulator (SLM). Several groups \cite{aulbach,katz,mccabe} have recently extended this approach to the time domain and demontrated the spatio-temporal focusing of femtosecond pulses through highly scattering media.

A general treatment of such linear complex media is provided by the scattering matrix formalism, as defined in \cite{beenakker} for instance. For ultrasound or microwaves, such a formalism is particularly appropriate as the wave field can be generated and detected by arrays of independent elements. Popoff \textit{et al.} \cite{popoff2010measuring,popoff2010image} recently extended this method to optics and measured the transmission matrix for light between a SLM (emission) and a CCD camera (reception) through a scattering medium. The latter is a subpart of the scattering matrix linking the forward scattered field to the incident field. The experimental access to this transmission matrix is particularly important since it contains many informations on the propagation medium. In particular, it should give access to the open channels that allow to transmit light through the scattering sample~\cite{pendry,beenakker,vellekoop2}.

In this Letter, we study another subpart of the scattering matrix: the backscattering matrix $\mathbf{K}$ (BM). The knowledge of the BM brings more fundamental insight into the medium. Initially developed for ultrasound~\cite{prada1994time,prada2,FinkPhysTo97} and then extended to microwaves~\cite{tortel,micolau2003dort}, the DORT method (French acronym for Decomposition of the Time Reversal Operator) takes advantage of the BM to focus iteratively by time reversal processing on each scatterer  of a multi-target medium.  Derived from the theoretical study of the iterative time reversal process, this approach relies on the diagonalization of the time reversal operator $\mathbf{KK}^{\dag}$. A one-to-one association between each eigenstate and each scatterer allows focusing on each scatterer by transmitting the computed eigenvectors. Technically, a singular value decomposition (SVD) of the BM is used to compute the eigenstate of the time reversal operator. In this letter, we will demonstrate the DORT method in the optical regime and show its efficiency in weakly scattering media. An experimental setup is proposed to record the BM in optics. Using the DORT method, we perform selective focusing on 100 nm gold nanoparticles through an aberrating medium, opening an original alternative to conventional adaptive optics~\cite{tyson1991principles}. In a second part, we show how the DORT method can be used to discriminate and selectively excite the different scattering patterns of a single nanoparticle, hence paving the way towards a better characterization, imaging and control of nano-objects for instance in plasmonics~\cite{gjonaj_active_2011} or for super-resolution~\cite{PhysRevLett.102.213905}.

\begin{figure}[ht]
\center
\includegraphics[width=8.5cm]{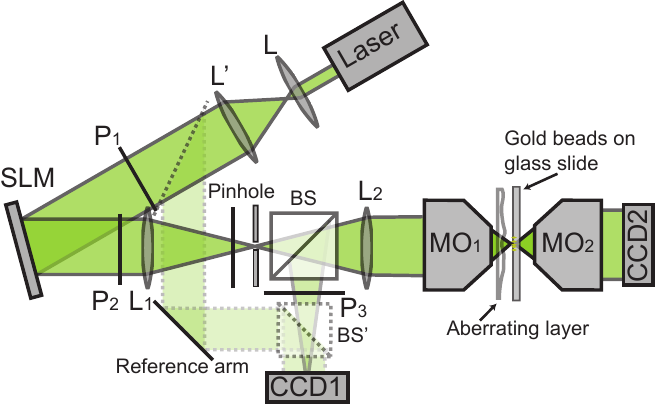}
\caption{Schematic of the apparatus. The 532 nm laser is expanded and reflected off a Holoeye LC-R 2500 liquid cristal SLM. Polarization optics (polarizer P$_1$ and analyser P$_2$) allow phase modulation. The modulated beam is focused on the plane containing the gold particules by a 20x objective (NA = 0.5) and the backscattered intensity is imaged by a first CCD-camera (CCD1). The second CCD-camera (CCD2) images the same plane in transmission with a 60x oil immersion objective (NA = 1.4). Analysers P$_2$ and P$_3$ are oriented  in a cross-polarized configuration. L, lens. BS, beamsplitter.}
\label{Setup}
\end{figure}

The experimental setup used for the measurement of the BM is displayed in Fig.~\ref{Setup}. To generate the incident wavefront, the beam from a single mode laser at 532 nm is expanded and spatially modulated by a liquid crystal SLM. Using a polarizer and an analyzer, we choose an appropriate combination of polarizations before and after the SLM to achieve a $2\pi$-phase modulation with a residual amplitude modulation below 15$\%$. Light is focused onto the surface of a glass slide on which is deposited a dilute solution of 100 nm isotropic gold beads. We image the output backscattered field pattern onto a first CCD camera (CCD1) thanks to a beamsplitter. To minimize specular reflections, the two polarizers are in a cross-polarized configuration. A reference arm can be used to make the backscattered field interefere with a plane wave onto CCD1. The system is set such that a pixel of the SLM corresponds approximately to an incident \textbf{k}-vector. An aberrating slab is placed 0.5 mm from the image plane between the scattering sample and the objective. It consists in an irregular PDMS polymer layer with a mean thickness of 200 $\mu$m deposited on a cover glass. The CCD1 records the image of the intensity in the plane containing the gold particles, with and without the aberrator. For control purposes, another CCD camera (CCD2) images the same plane in transmission, through a high NA objective.

In the experimental setup shown in Fig.~\ref{Setup}, the input and output modes correspond to  SLM and CCD1 macropixels, respectively. The BM corresponds to a matrix $\mathbf{K}$ whose complex coefficients $k_{mn}$ connect the optical response (in amplitude and phase) between the $n^{th}$ SLM pixel to the $m^{th}$ CCD1 pixel. In other words, the complex optical field on the $m^{th}$ outgoing pixel reads $E^{out}_m = \sum_n {k_{mn}E^{in}_n}$ with $E^{in}_n$ the field on the $n^{th}$ input pixel. $\mathbf{K}$ is of dimension $M \times N$, with $M$ and $N$ the number of pixels contained by CCD1 and the SLM, respectively. 
The inter-element responses $k_{mn}$ are measured following the procedure described in~\cite{popoff2010measuring}, \textit{i.e} using phase-shifting interferometry and a Hadamard input basis.

A singular value decomposition (SVD) of the BM  consists in writing $\mathbf{K} = \mathbf{U\Sigma V^{\dag}}$ where the subscript $^{\dag}$ denotes the transpose conjugate. $\mathbf{\Lambda}$ is a diagonal matrix containing the real positive singular values $\lambda_{i}$ in a decreasing order $\lambda_1>\lambda_2> \cdots >\lambda_N$. $\mathbf{U}$ and $\mathbf{V}$ are unitary matrices whose columns correspond to the output and input singular vectors $\mathbf{U_i}$ and $\mathbf{V_i}$, respectively. The DORT method consists in using the SVD of the BM to  efficiently detect and separate the responses of several scatterers in homogeneous or weakly heterogeneous media \cite{prada2}. Indeed, under the single scattering approximation and for pointlike scatterers, each scatterer is associated mainly to one significant eigenstate. The latter is linked to a non-zero singular value $\lambda_i$ that relates to the scatterer reflectivity. In first approximation, each singular vector $\mathbf{V_i}$ of $\mathbf{K}$ corresponds to the complex amplitude mask which, when applied to the SLM, focuses onto the associated scatterer. 
\begin{figure}[ht]
\center
\includegraphics[width=8.5cm]{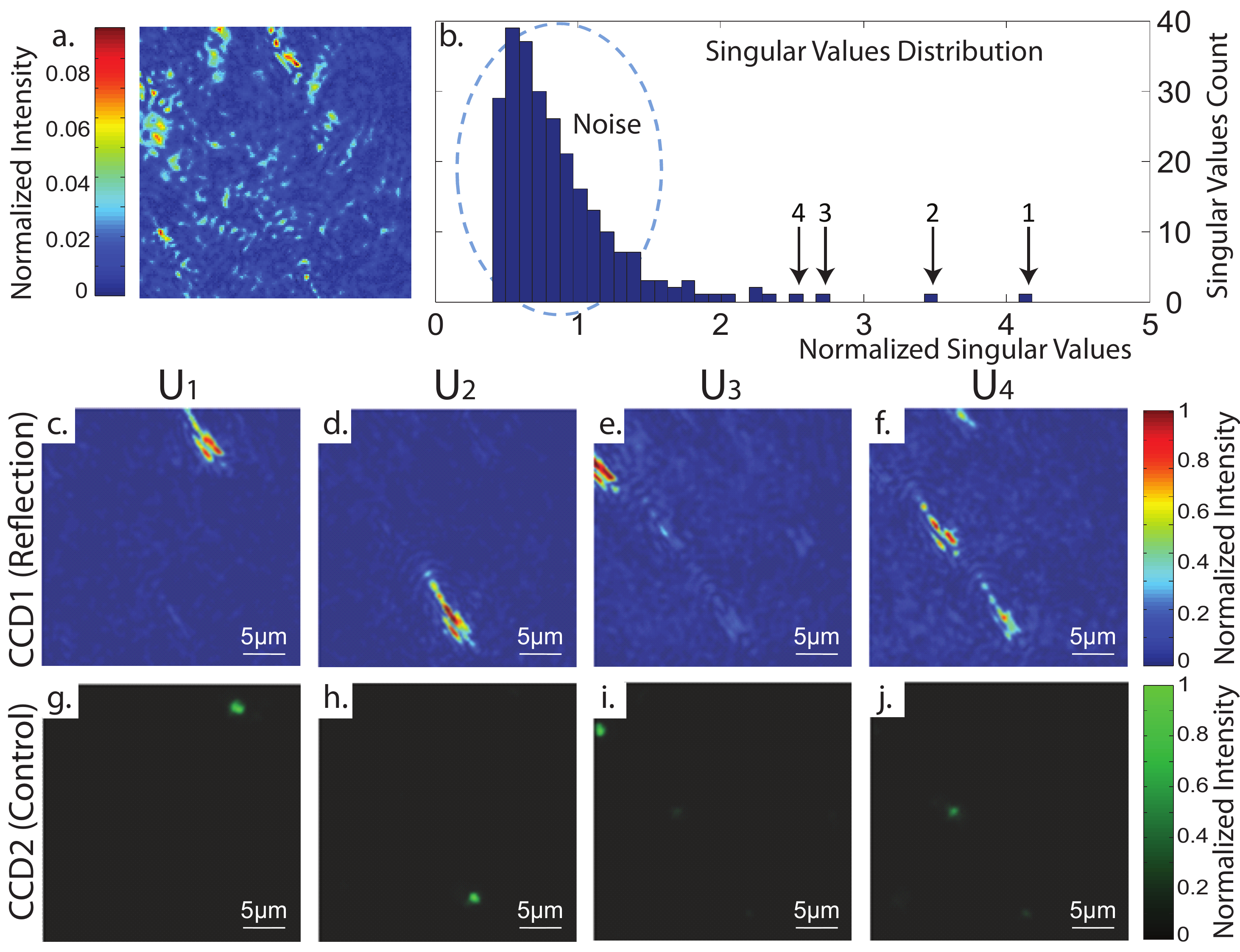}
\caption{ Experimental results for selective focusing on nanobeads  through an aberrating layer using the DORT method. \textbf{a.} aberrated image of the object plane on CCD1 in reflection for a plane wave illumination \textbf{b.}, singular value distribution of the measured BM. \textbf{c.-f.} ( resp. \textbf{g.-j.}) ; images  in reflection by camera CCD1 ( resp. in transmission by the control camera CCD2), when sending the four first singular vectors.  All images from CCD1 (resp. CCD2) share the same intensity scale.}
\label{MultTarg}
\end{figure}

\begin{figure}[ht]
\center
\includegraphics[width=8.5cm]{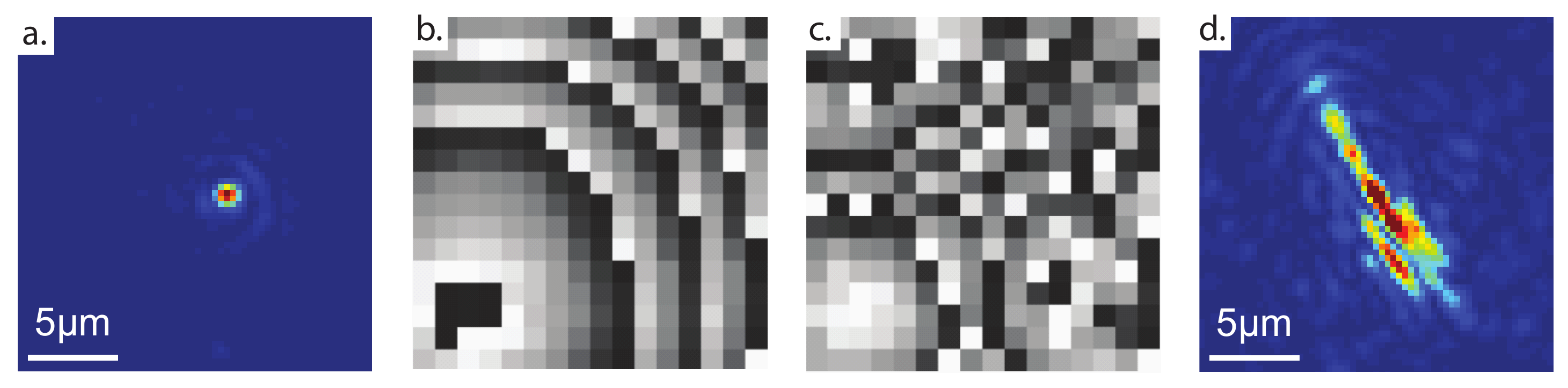}
\caption{Phase of the first input singular vector $\mathbf{V_1}$ (b, c) and the corresponding focal spots recorded by CCD1 (a, d) in free space and in presence of an aberrating layer, respectively. }
\label{InputWF}
\end{figure}

Figs.~\ref{MultTarg} and ~\ref{InputWF} show the experimental results of the DORT method applied to the BM matrix when several 100 nm gold beads are dispersed on the slide placed behind an aberrating layer. For this experiment, we have used the field generated by the unmodulated part of the illumination as the reference for the interferometric measurement rather than the plane wave from the reference arm. This method detailed in~\cite{popoff2010measuring} allows for a better stability without being detrimental to focusing applications. The numbers of pixels have been fixed to $N=M=1024$. Fig.~\ref{MultTarg}(a) shows the image measured by CCD1 in reflection when a plane wave is sent from the SLM. The image of the medium is drastically damaged by the aberrating layer: the number of gold beads as well as their position and reflectivity cannot be deduced from this image. In contrast, the DORT method gives an unambiguous answer to these questions. The singular value distribution (Fig.~\ref{MultTarg}(b)) of the BM exhibits two parts. The continuum of low singular values is associated to the noise in the BM (residual specular reflections, laser fluctuations, CCD readout noise \textit{etc.}). A few higher singular values are associated to the brightest gold nanoparticles since the singular value is directly proportional to the reflectivity of the scatterer. The input singular vector $\mathbf{V_i}$ corresponds to the wave-front focusing on the $i^{th}$ particle and the output singular vector $\mathbf{U_i}$ is the corresponding scattering pattern measured by CCD1 in reflection (see Figs.~\ref{MultTarg}(c,d,e,f)). Figs.~\ref{MultTarg}(g,h,i,j) correspond to the control images measured in transmission by CCD2, when the first four singular vectors $\mathbf{V_i}$ are back-propagated from the SLM. These images show the selective focusing performed by DORT: each singular vector back-propagates selectively on one gold bead without illuminating the other ones. Quantitatively, an intensity enhancement of $\sim$ 20 is observed on the four gold beads position when DORT focusing is performed compared to plane wave illumination. 

Fig.~\ref{InputWF} shows explicitly the effect of aberrations by comparing the singular vectors $\mathbf{V_1}$ and $\mathbf{U_1}$ (associated to the brightest scatterer) with and without the aberrating layer. Figs.~\ref{InputWF}(b) and (c) display the phase of $\mathbf{V_1}$ on the SLM in both situations. In free space, the wavefront focusing on a particle corresponds to a Fresnel phase zone plates (Fig.~\ref{InputWF}(b)). In contrast, in presence of an aberrating layer (Fig.~\ref{InputWF}(c)), $\mathbf{V_1}$ is strongly modified to compensate for the wavefront distortions. Figs.~\ref{InputWF}(a) and (d) display a zoom on the corresponding focal spots recorded by CCD1. The comparison of both figures shows the damaging effect of the aberrating layer and highlights the benefits of the DORT method in this configuration. 

So far, we made the assumption that each scatterer is only associated to one single singular value. However, this is not strictly true: there can be several singular values associated to a single scatterer, each one corresponding to a different scattering pattern of the particle~\cite{chambers,minonzio2005characterization,minonzio2008characterization}. To investigate the different eigenstates associated to one nanoparticle, we now study analytically and experimentally the DORT analysis of the BM without any aberrating layer and with only one gold particle in the observation area.

Let us assume that the incident field is linearly polarized along $x$, $z$ is the propagation axis and $y$ the polarization observed in reflection (cross-polarized configuration). The BM can be decomposed into the product of three matrices: $\mathbf{K} = \mathbf{K^{\prime\prime}.\bar{\alpha}.K^{\prime}}$.
Here $\mathbf{K^{\prime}}$ is a  $3 \times N$ matrix describing the propagation of light from the $N$ pixels of the SLM to the scatterer. The three lines of $\mathbf{K^{\prime}}$ correspond to the different orientations of the electric field at the particle position.  $\mathbf{\bar{\alpha}}$ is the polarisability tensor of the particle linking the local electric field to the dipole moment $\mathbf{p}$ of the particle. $\mathbf{K^{\prime\prime}}$ is a $M \times 3$ matrix which connects the dipole moment $\mathbf{p}$ to the $y$-component of the electric field measured on the $M$ pixels of CCD1. We will assume that the particle is isotropic: $\mathbf{\bar{\alpha}} = \alpha \mathbf{I}$ with $\alpha$ the particle polarizability and $\mathbf{I}$ the identity matrix.

\begin{figure}[t]
\center
\includegraphics[width=8.5cm]{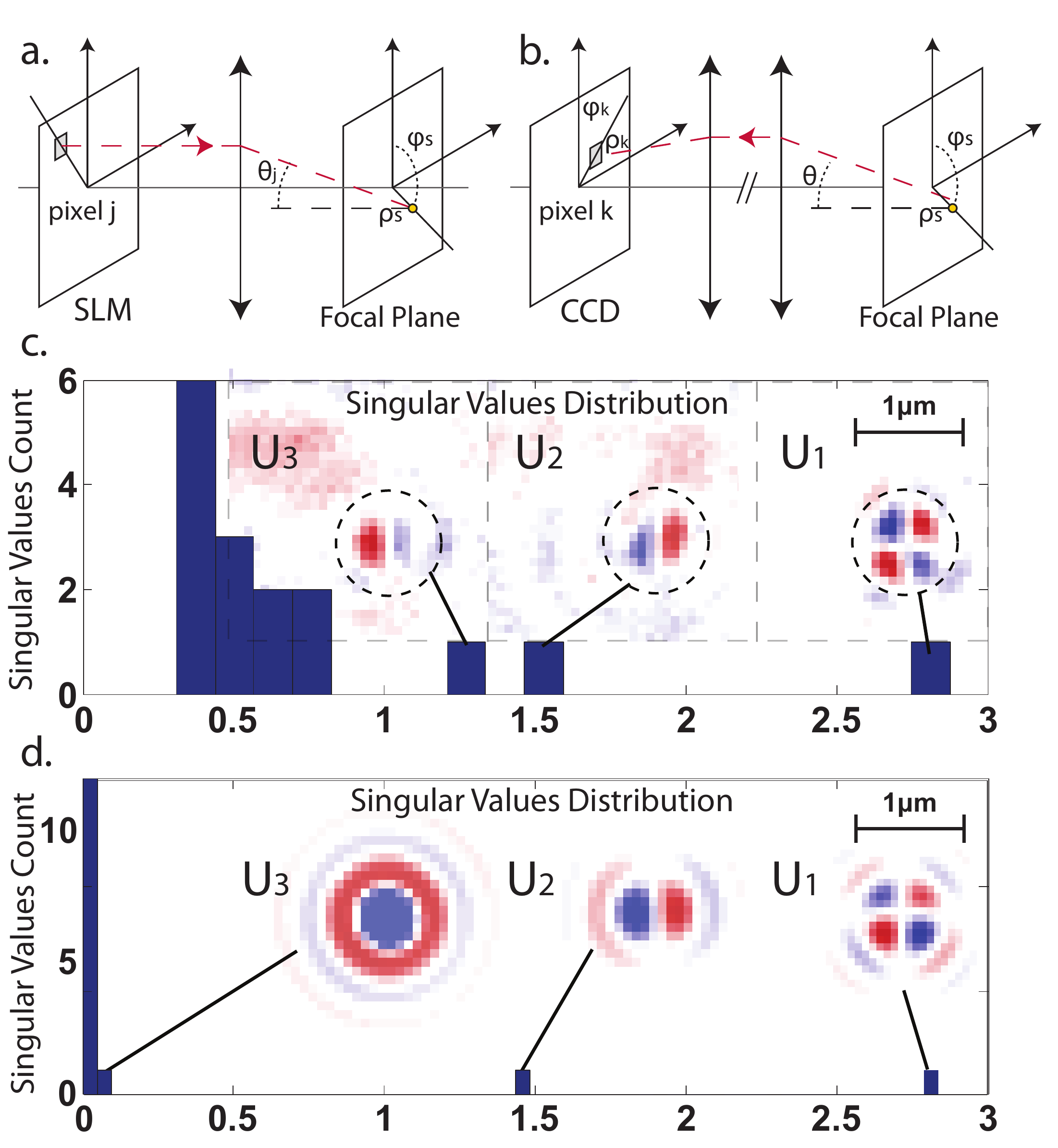}
\caption{Singular value decomposition of the experimental and theoretical BM with a single scatterer. (a) and (b) represent the emission and the recording geometry of the experiment.(c) and (d) show the singular value distributions of the measured and calculated BM. The three first output singular vectors are shown in the insets.}
\label{SingleScat}
\end{figure}

The electric field illuminating the particle in the focal plane can be deduced from~\cite{wilson1997imaging,richards1959electromagnetic} and leads to the following expression for the coefficients of $\mathbf{K^{\prime}}$:
\begin{eqnarray}
\label{K'}
k^{\prime}_{1j} & =& I_0(\theta^1_j,\rho_s) + \underbrace{I_2(\theta^1_j,\rho_s)\cos(2\phi_s)}_{negligible} \\
\label{K'2}
k^{\prime}_{2j} & =& I_2(\theta^1_j,\rho_s)\sin(2\phi_s) \\
\label{K'3}
k^{\prime}_{3j} & =& 2jI_1(\theta^1_j,\rho_s)\cos(\phi_s)
\end{eqnarray}
$\phi_s$ and $\rho_s$ are the polar coordinates of the particle in the focal plane (see Fig.~\ref{SingleScat}(a)). $\theta^1_j < \theta_1^{max}$ is the angle made by the \textbf{k}-vector associated to the $j^{th}$ element of the SLM with the optical axis. The functions $I_0$, $I_1$ and $I_2$ are given by~:
\begin{eqnarray}
I_0(\theta,\rho) &=& \sqrt{\cos\theta}\sin\theta(1+\cos\theta)J_0(2\pi \rho\sin\theta)\\ 
I_1(\theta,\rho) &=& \sqrt{\cos\theta}\sin^2 \theta J_1(2\pi \rho\sin\theta)\\
I_2(\theta,\rho) &=& \sqrt{\cos\theta}\sin\theta(1-\cos\theta)J_2(2\pi \rho\sin\theta)
\end{eqnarray}
with $J_0$, $J_1$ and $J_2$ the zero, first and second order spherical Bessel functions of the first kind, respectively.

The $y$-component of the backscattered field can be also deduced from the calculations of~\cite{wilson1997imaging} and leads to the following expression for $\mathbf{K^{\prime\prime}}$:
\begin{eqnarray}
\label{kpp1}
k^{\prime\prime}_{k1}& = & \int_0^{\theta^{max}_2} \, \,  {d\theta_2 I_2(\theta_2,\rho_k)}\sin(2\phi_k)\\
\label{kpp2}
k^{\prime\prime}_{k2}& = & \int_0^{\theta^{max}_2} {d\theta_2  [ I_0(\theta_2,\rho_k) +  \underbrace{I_2(\theta_2,\rho_k)\cos(2\phi_k) }_{negligible}]} 
\\ 
\label{kpp3}
k^{\prime\prime}_{k3}& = & 2j\int_0^{\theta^{max}_2} d\theta_2 {I_1(\theta_2,\rho_k)}\sin(\phi_k)
\end{eqnarray}
$\rho_k$ and $\phi_k$ are the polar coordinates of the $k^{th}$ pixel of CCD1 and $\theta_2$ is the angle made by a beam diffracted by the particle with the optical axis before going through the microscope objective (see Fig.~\ref{SingleScat}(b)). Hence, the BM depends on the following parameters: the position of the particle in the focal plane ($\phi_s$,$\rho_s$), the numbers of SLM and CCD elements ($N$ and $M$) and the input and output maximum angles ($\theta^{max}_1$ and $\theta^{max}_2$). $\theta^{max}_2$ is directly given by the numerical aperture of the optical system, but $\theta^{max}_1$ can be smaller if the area of illumination is smaller than the area of the pupil of the microscope objective. 
Theory predicts that the $x$, $y$ and $z$ components of the particle dipole moment $\mathbf{p}$ give rise to different scattering patterns:  monopolar for $y$ (Eq.\ref{kpp2}), bipolar for $z$ (Eq.\ref{kpp3}) and quadrupolar for $x$ (Eq.\ref{kpp1}) (see Fig.\ref{SingleScat}(d)). Those scattering patterns are strictly orthogonal if they are observed over a solid angle $\Omega=2\pi$. Hence, in theory, the output singular vectors $\mathbf{U_i}$ of $\mathbf{K}$ should correspond to the different scattering patterns of the nanoparticle (\textit{i.e} the columns of $\mathbf{K^{\prime\prime}}$). In practice, this is not strictly true since only a limited part of the half space is accessible in our experiment.

Our theoretical predictions are now confronted to the experimental results. As an exact measurement of the output singular vectors is needed, the complex wave field has to be measured directly. A reference arm is now used to make the backscattered wave interfere with a reference plane wave on CCD1 (see Fig.\ref{Setup}). The output singular vectors $\mathbf{U_i}$ correspond then to the complex scattering patterns of the particle. The number $N$ of controlled segments on the SLM has been lowered to $16$ for stability contraints over the time of the experiment. The result of the DORT analysis is shown in Fig.\ref{SingleScat}. The experimental and theoretical singular values distribution as well as the three first output singular vectors of the BM are compared in Fig.\ref{SingleScat}(c) and (d), respectively. The experimental values of $\theta^{max}_1$ and $\theta^{max}_2$ have been used for our theoretical calculations ($\theta^{max}_1=22 °$, $\theta^{max}_2=68°$ ). The position of the particle in the focal plane has been arbitrarily fixed ($\phi_s=7/8\pi$, $\rho_s=665$ $nm$) such that the theoretical and experimental ratios between the first and the second singular values are equal.Our theoretical calculations predict that only the three first singular values are non-zero, each one corresponding to the independent scattering patterns of the three orthogonal dipolar radiations (along $x$, $z$ and $y$ resp.). The experimental results display a continuum of low singular values corresponding to noise and three higher singular values (Fig.\ref{SingleScat}(d)). The two first output singular vectors match the scattering patterns of the $x-$ and $z-$components of the nanoparticle dipole moment, as predicted theoretically. The third output singular vector has a geometry comparable to the second one, but is more affected by the experimental noise since a significant part of its energy is observed outside the particle. This singular vector might correspond to a coupling between the second singular state ($z-$component) and  eigenmodes of the noise background. The scattering pattern associated to the $y-$component of the dipole moment is drowned in noise and could not be observed.

The DORT method allows to excite independently each component of the nanoparticle dipole moment: such an experiment would require to increase the number of SLM segments for quantitative results and a near field scanning optical microscope (NSOM) to probe the near-field of the nanoparticle. For larger scatterers, the singular spectrum may be even richer: more radiation modes would be excited and a better characterization of the nanoparticle could be obtained (size, position, permittivity \textit{etc.}). In particular, one could discriminate the quadrupolar radiation of the particle, which may open important perspectives in plasmonics. 

In summary, we have performed an experimental measurement of the BM of a weakly scattering medium. We have used this BM to demonstrate the DORT method in optics and proved that it allows selective focusing on nanoparticles and their caracterization. This approach could lead to many applications in nano-optics, sensing or imaging and entails the interest of the BM in optics. We are confident that measuring the BM in more complex media will raise interesting problems in optics. For instance, the measurement of the BM in highly scattering media may allow the separation of the single and multiple scattering contributions \cite{aubry2009random} as well as the study of the coherent backscattering cone and of the diffusive halo \cite{aubry2}. 

We wish to thank Claire Prada and Sébastien Bidault for inspiring discussions as well as Laurent Boitard and Gilles Tessier for their help in the preparation of the samples. This work was made possible by financial support from ''Direction Générale de l'Armement'' (DGA), Agence Nationale de la Recherche via grant ANR 2010 JCJC ROCOCO, Programme Emergence 2010 from the City of Paris and BQR from Université Pierre et Marie Curie and ESPCI.

\end{document}